\documentstyle{article}
\newcommand{\bvec}[1]{\mbox{\boldmath ${#1}$}}
\newcommand{\Shyp}[2]{$^{#1}_{\Sigma}${#2}}
\newcommand{\Lhyp}[2]{$^{#1}_{\Lambda}${#2}}
\newcommand{\Yhyp}[2]{$^{#1}_{Y}${#2}}
\begin{document}
 \title{Electroproductions of Light $\Lambda$- and $\Sigma$-Hypernuclei}
 \author{Shoji Shinmura\\Department of Applied Mathematics\\
         Faculty of Engineering, Gifu University\\
         Yanagido 1-1, Gifu, Japan}
 \date{January 23, 1994}
 \maketitle
\begin{abstract}
  Theoretical estimations of production cross sections of
 light $\Lambda$ and $\Sigma$ hypernuclei in $(e,e'K^{+})$
 reactions at around CEBAF energies are given.
 Because of dominant spin-flip amplitudes and
 large momentum transfers,
 unnatural parity states and stretched states of hypernuclei
 are favorably excited. They are compared with quasifree hyperon
 productions.
\end{abstract}
\section{Introduction}
 In last two decades, hypernuclear physics is remarkably developed
 both in theoretical and experimental aspects.
 The $(K^{-},\pi)$ and $(\pi,K^{+})$
 experiments have been performed intensively and clarified properties
 of many hypernuclear states.  However, such states are only a part
 of possible hypernuclear states. For example, ground states and
 deep-hole states of heavy hypernuclei are
 hardly excited in these experiments because of strong absorption of
 $\pi$ and $K^{-}$ in nuclear medium. Unnatural parity states are
 also weakly excited because of the small spin-flip amplitudes.
\par
 As alternative tools to produce hypernuclei, photo- and
 electro-productions of hypernuclei, that is, hypernuclear productions
 in the $(\gamma,K^{+})$ and $(e,e'K^{+})$ reactions, are discussed.
 These reactions contain only particles which interact weakly with
 nuclear medium.
 Therefore, these are favorable to produce directly hypernuclei with
 a deeply bound $\Lambda$ and/or
 a nucleon deep-hole. Further, the dominant
 spin-flip amplitudes in these reactions are of great advantage to
 excite unnatural parity states of hypernuclei.
\par
  Theoretical estimations\cite{Hsi}-\cite{Sot} of photoproductions of
 hypernuclei have been performed by several authors. But the
 electroproductions\cite{Hsi} were treated under rough approximations
 or only for the $\Lambda$-hypernuclei\cite{Coh}.
  In the present work, to improve such a situation, we give
 theoretical estimations of the electroproduction cross sections both
 for $\Lambda$ and $\Sigma$-hypernuclei.
 In spite of uncertainties in the elementary reactions,
 we treat the energy region corresponding to CEBAF, which gives
 high possibility to the experiment in near future.
\par
  Our calculations are based on a relativistic impulse approximation and
 the Walecka model of nuclei and hypernuclei.
 The elementary reactions are treated relativistically without any
 approximation within the model.  Parameters in the model are
 determined to reproduce known experimental data.
\par
 In \S2, we give the model of elementary processes. In \S3, the
 relativistic model for nuclei and hypernuclei are explained.
 In \S4, the relativistic impulse approximation for hypernuclear
 production is formulated. Numerical results are given in \S5.
 The results are compared with the quasifree hyperon production which
 is the largest background process in the experiments in \S6.
 Conclusions of this work is given in \S7.

\section{Elementary processes $N(e,e'K^{+})Y(Y=\Lambda,\Sigma)$}
  Theoretical models of processes including the strong
 interaction cannot be determined uniquely. We can only pick up
 possible mechanisms and determine the parameters in the model so as
 to reproduce known empirical data.
  We assume that the processes $N(e,e'K^{+})Y$ consist of
 one-photon-exchange and $N(\gamma,K^{+})Y$ vertex, as shown Fig.1a.
  The former is
 theoretically undoubted  because of the small QED coupling constant.
 The latter is complicated and is understood only insufficiently.
\par
  As a model of the $N(\gamma,K^{+})Y$ vertex, we employ a sum of
 one-particle-exchange mechanisms as Fig.1b. This model is commonly
 used in theoretical studies\cite{Tho}-\cite{Ad3}.
 The coupling constants in the model are determined by fitting
 the experimental data. In the present work, we use the values in
 refs.\cite{Hsi} and \cite{Be2}, which are listed in Table 1
 and 2 for $Y=\Lambda$ and $\Sigma$, respectively.
 It should be noted that the model and the values of parameters have
 large uncertainties because of the limitation of experimental data.
\par
 Details of calculations with these parameters are given
 in refs.\cite{Hsi} and \cite{Be2}.
 To check our computer code, we repeated their calculations.
 As a result, we obtained alomost the same results with those in
 the original works. Only one exception is an oppsite assignment of
 lines in the Fig.2 of ref.\cite{Hsi}.
\par
  Using the model, we calculate the elementary processes
 $N(e,e'K^{+})Y$ at around CEBAF energies. As expected,
 we find vary large spin-flip amplitudes in all cases.
\par
  As a comparison, we can also use the approximate expression,
\begin{equation}
 \frac{d^{3}\sigma}{d\Omega_{e'}dE_{e'}d\Omega_{K}} =
    \frac{1}{2\pi^{2}}\frac{\alpha}{1-\epsilon}
    \frac{E_{e'}}{E_{e}}\frac{k_{\gamma}}{(E_{\gamma}^2-k_{\gamma}^2)}
    \frac{d\sigma}{d\Omega_{K}}(\gamma,K),
\end{equation}
 which is used in ref.\cite{Hsi}. In this expression, only on-shell
 $N(\gamma,K^{+})Y$ amplitudes are assumed.
 But this gives
 different results from our exact calculations by around 20-40\%.
\par
 It should be noted that the present model give large cross sections
 for $n(e,e'K^{+})\Sigma^{-}$
 in comparison with those for $p(e,e'K^{+})\Sigma^{0}$
 by about one order of magnitude. Therefore, the $\Sigma$
 hypernuclear productions are sensitive to neutron wave functions in
 the target nucleus
\begin{table}
 \caption{Coupling constants$^{a)}$ for $\gamma+p\rightarrow K^{+}+\Lambda$.}
 \begin{center}
 \begin{tabular}{cccc} \hline
  $\frac{g_{\Lambda}}{\sqrt{4\pi}}$ & $\frac{G_{\Sigma}}{\sqrt{4\pi}}$
 & $\frac{G_{V}}{4\pi}$ & $\frac{G_{T}}{4\pi}$  \\ \hline
  2.57 & 1.52 & 0.105 & 0.064 \\ \hline
 \multicolumn{4}{l}{a)These are assigned by set1 in ref.\cite{Hsi}.}\\
 \end{tabular}
 \end{center}
\end{table}
\begin{table}
 \caption{Coupling constants$^{a)}$ for $\gamma+N\rightarrow K^{+}+\Sigma$.}
 \begin{center}
 \begin{tabular}{lcccccc} \hline
  & $\frac{g_{\Sigma}}{\sqrt{4\pi}}$ & $\frac{G_{\Lambda}}{\sqrt{4\pi}}$
  & $\frac{G_{V}}{4\pi}$&$\frac{G_{T}}{4\pi}$&$\frac{G^{1}_{\Delta}}{4\pi}$
  & $\frac{G^{2}_{\Delta}}{4\pi}$ \\ \hline
 PS       &  2.20 & -4.82 & 0.113 & -0.038 &   0    & 0     \\
 PS+delta &  2.72 & -3.60 & 0.104 &  0.005 & -0.069 & 0.314 \\ \hline
 \multicolumn{7}{l}{a)These are taken from ref.\cite{Be2}} \\
 \end{tabular}
 \end{center}
\end{table}
\begin{table}
 \caption{Potentials for nucleon and calculated single
          particle energies for protons(in MeV and fm)}
 \begin{center}
 \begin{tabular}{lccc} \hline
            &  {$^{4}$He} & {$^{12}$C} & {$^{16}$O}  \\ \hline
  $S_{0}$   &  -286.0     &   -288.0   &   -426.0    \\
  $V_{0}$   &   200.0     &    215.0   &    340.0    \\
  $a$       &     -       &      0.7   &      0.7    \\
  $r_{0}$   &    1.2      &      1.2   &    1.055    \\ \hline
  $s_{1/2}$ &  -20.1      &    -34.9   &   -40.3     \\
  $p_{3/2}$ &   -         &    -15.3   &   -18.3     \\
  $p_{1/2}$ &   -         &      -     &   -12.0     \\ \hline
 \end{tabular}
 \end{center}
\end{table}
\begin{table}
 \caption{Potentials for hyperons and calculated single particle
 energies(in MeV and fm)}
 \begin{center}
 \begin{tabular}{lccccccc} \hline
     & \Lhyp{4}{H} & \Lhyp{4}{H}$^{*}$ & \Lhyp{16}{N} & \Lhyp{12}{B}
     & \Shyp{4}{H}(\Shyp{4}{H}$^{*}$) & \Shyp{16}{N} & \Shyp{12}{B} \\
 \hline
 $S_{0}$ & -142.5 & -137.2 & \multicolumn{2}{c}{-187.0} & -243.0
 & \multicolumn{2}{c}{-187.0} \\
 $V_{0}$ & 100.0  & 100.0  & \multicolumn{2}{c}{155.0} & 200.0
 & \multicolumn{2}{c}{177.0} \\
 $a$     & -  & -  & \multicolumn{2}{c}{0.7} &
 & \multicolumn{2}{c}{0.7} \\
 $r_{0}$ & 1.0 & 1.0 & \multicolumn{2}{c}{1.15} & 1.0
 & \multicolumn{2}{c}{1.15} \\ \hline
  $s_{1/2}$ &-2.1 & -1.0 & -13.3$^{a)}$ & -11.6 & -3.0 &-5.1 &-3.5 \\
  $p_{3/2}$ & -   &  -   & -2.7$^{a)}$  & -1.1  & -    & -   &  -  \\
  $p_{1/2}$ & -   &  -   & -1.7$^{a)}$  & -0.3  & -    & -   &  -  \\
 \hline
  \multicolumn{8}{l}{a)for \Lhyp{15}{N}(see text)} \\
 \end{tabular}
 \end{center}
\end{table}

\section{Relativistic model of nuclei and hypernuclei}

 In the present work, we treat relativistically elementary reactions
 and hypernuclear productions.
 For consistency, we should employ a
 relativistic model of nuclei and hypernuclei.
 The most famous is the Walecka model. In this model, nucleons and
 hyperons are assumed to behave as Dirac particles and are moving in
 attractive scalar potential and repulsive vector potential in
 nuclear medium.
 The Dirac equation for nucleon(hyperon) is given by
\begin{equation}
  [i\partial\!\!\!/ - m - S(r) - \gamma_{0}V(r) ]\psi(\bvec{r})e^{-iEt}
                                                                   = 0,
\end{equation}
  where, $m$ is the reduced mass between nucleon(hyperon) and core
 nucleus. $S(r)$ and $V(r)$ are
the scalar and vector potentials for nucleon (hyperon) in nuclei
(hypernuclei).
  This equation is solved numerically and the single particle energy
 is given by $E-m$.
\par
  The potentials $S(r)$ and $V(r)$ are determined so as to reproduce
 phenomenological single particle energies(SPE) in nuclei and
 hypernuclei. In this work, except for the $A=4$ cases, the potentials
 are assumed to have the Woods-Saxon shape like as
\begin{eqnarray}
       S(r) & = & S_{0}[1+\exp((r-R)/a)]^{-1}, \\
       V(r) & = & V_{0}[1+\exp((r-R)/a)]^{-1}.
\end{eqnarray}
 For $A=4$ cases, the Gaussian shape is assumed;
\begin{eqnarray}
       S(r) & = & S_{0}\exp(-(r/R)^{2}), \\
       V(r) & = & V_{0}\exp(-(r/R)^{2}).
\end{eqnarray}
 $S_{0}$,$V_{0}$, $R=A^{1/3}r_0$, and $a$ used in
 the present calculations are displayed in Table 3 and 4.
\par
  The potentials for nucleons are given in Table 3. In $^{16}$O(
 \Yhyp{16}{N})
 and $^{12}$C(\Yhyp{12}{B}), the potentials are determined to give
 phenomenological proton SPE in $s_{1/2}$, $p_{3/2}$ and
 $p_{1/2}$(only for
 $^{16}$O(\Yhyp{16}{N})) states.
 On the other hand, those in $^{4}$He(\Yhyp{4}{H}) are determined so
 as to reproduce the binding energy and the root-mean-square radius
 of $^{4}$He, simultaneously.
\par
  For hyperons, the potentials are given in Table 4. Those for the
 $\Lambda$
 particle give phenomenological SPE in \Lhyp{15}{N}(those in
 \Lhyp{16}{N} are
 not available) and \Lhyp{12}{B}.
 We can easily understood weaker strengths of
 $S$ and $V$ than those for nucleon by the small spin-orbit splitting
 for the $\Lambda$ particle.
 The potentials in \Lhyp{4}{H} and \Lhyp{4}{H}$^{*}$ are
 determined to reproduce their experimental $\Lambda$ separation
 energies, respectively.
\par
  For $\Sigma$ particle, phenomenological informations are very limited.
 Therefore, we assume some trial values.
 In \Shyp{4}{H}, the potential for $\Sigma$ are determined to
 give $B_{\Sigma}=3.0$MeV, which
 is near the experimental data\cite{Hay} and theoretically
 predicted values\cite{Ha1}-\cite{Ha3}.
 \Shyp{4}{H}$^{*}$ with spin=1 is a hypothetical
 object with the same binding energy with \Shyp{4}{H}.
 This is introduced in order to clarify features of the
 electroproductions.
 In \Shyp{16}{N} and \Shyp{12}{B}, the potentials
 give -5.1MeV and -3.5MeV for $s_{1/2}$ SPE, respectively. In
 both systems, all $p$-states are unbound.
\par
  It should be noted that the quantitative modifications of the
 potentials are not so sensitive to qualitative features of
 hypernuclear production discussed later, because they
 are mainly determined by the angular momentum and parity selections
 and high momentum components of wave functions, which is weakly
 influenced by the binging energies.

\section{Hypernuclear productions in Relativistic Impulse
 Approximation}

Since both the incident and outgoing particles have high energies
 compared with nucleon binding energies in nuclei and momentum
 transfers in the reaction are considerably larger than Fermi momenta
 of nuclei, the relativistic impulse approximation is expected to
 work well. In fact, this approximation is used successfully in
 various hypernuclear production reactions.
\par
In the relativistic impulse approximation, the hypernuclear
 production cross section in the $(e,e'K^{+})$ reaction is given by
\begin{eqnarray}
  d\sigma & = & (2\pi)^{4}\delta^{4}(p_{e}+p_{A}-p_{e'}-p_{B}-p_{K})
\frac{d^{3}\bvec{p}_{e'}m_e}{(2\pi)^{3}E_{e'}}
\frac{d^{3}\bvec{p}_{K}}{(2\pi)^{3}E_{K}}
\frac{d^{3}\bvec{p}_{B}}{(2\pi)^{3}}\\
& & \frac{(2m_{p})(2m_{e})}{[(p_{e}
 \cdot p_{A})^{2}-p_{e}^{2}p_{A}^{2}]^{1/2}}
\frac{1}{2J_{i}+1}\sum_{M_{i},M_{f}}\left|<\!\!J_{f}M_{f}T_{f}N_{f}|
   T|J_{i}M_{i}T_{i}N_{i}\!\!>\right|^{2}, \nonumber
\end{eqnarray}
where, $p_{A}$($p_{B}$) and $|J_{i}M_{i}T_{i}N_{i}\!\!>$
($|J_{f}M_{f}T_{f}N_{f}\!\!>$) are the momentum and the quantum state
of the initial nucleus(the final hypernucleus), respectively. The
$T$-matrix is given by the nuclear matrix elements and the
elementary process as follows;
\begin{equation}
 <\!\!J_{f}M_{f}T_{f}N_{f}|T|J_{i}M_{i}T_{i}N_{i}\!\!>
    = \sum_{\alpha\alpha'}<\!\!J_{f}M_{f}T_{f}N_{f}|
     C_{\alpha'}^{\dagger}C_{\alpha}|J_{i}M_{i}T_{i}N_{i}\!\!>
     <\!\!\alpha'|t|\alpha\!\!>,
\end{equation}
where, $\alpha$ is a set of single particle quantum numbers,
$\{n,j,m;t,m_{t}\}$. In our simple model without configuration
 mixings, the nuclear matrix element given by
\begin{eqnarray}
 & & <\!\!J_{f}M_{f}T_{f}N_{f}|C_{\alpha'}^{\dagger}C_{\alpha}
|J_{i}M_{i}T_{i}N_{i}\!\!>\: = \sum_{JMTN}
    (-1)^{j-m+1/2-m_{t}+J_{i}-M_{i}+T_{i}-N_{i}} \nonumber \\
& & \hat{J}^{2}\hat{T}^{2}
     \left[\begin{array}{ccc}
             \!\!\! J_{f} & \!\!\! J_{i} & \!\!\! J \\
             \!\!\!-M_{f} & \!\!\! M_{i} & \!\!\! M
           \end{array}\right]\!
     \left[\begin{array}{ccc}
             \!\!\! T_{f} & \!\!\! T_{i} & \!\!\! T \\
             \!\!\!-N_{f} & \!\!\! N_{i} & \!\!\! N
           \end{array}\right]\!
     \left[\begin{array}{ccc}
             \!\!\! j'    & \!\!\! j     & \!\!\! J \\
             \!\!\! -m'   & \!\!\! m     & \!\!\! M
           \end{array}\right]\!
     \left[\begin{array}{ccc}
             \!\!\! t_{y} & \!\!\! 1/2   & \!\!\! T \\
             \!\!\! -m_{y}& \!\!\! m_{t} & \!\!\! N
           \end{array}\right],
\end{eqnarray}
for a definitely occupied nucleon state($\alpha=\{n,j,m,1/2,m_{t}\}$)
 in the initial nuclear state and a hyperon
 state($\alpha'=\{n',j',m',t_{y},m_{y}\}$) in the final hypernuclear
 state.
\par
  The elementary process in nuclear medium is given by
\begin{equation}
  <\!\!\alpha'|t|\alpha\!\!>\: = \int d^{3}\bvec{p}d^{3}\bvec{q}'
             \overline{\psi}_{\alpha'}(\bvec{p}')
             \phi_{K^{+}}^{(-)*}(\bvec{p}_{K},\bvec{q}')
             t(p_{e},p;p_{e'},p',q')
             \psi_{\alpha}(\bvec{p}).
\end{equation}
\par
In the present work, we use the plane-wave
 ($\delta^{3}(\bvec{p}_{K}-\bvec{q}')$)
instead of the distorted wave
($\phi_{K^{+}}^{(-)*}(\bvec{p}_{K},\bvec{q}')$)
 for outgoing kaon, because kaons with energies near 1 GeV does
 not strongly interact with nuclei and because
 only light nuclear targets are considered. The effects of the
 distortion of kaons by nuclear interaction were discussed in the
 photoproductions by several authors\cite{Be1}\cite{Sot}. The authors
 showed that the effects reduce the cross sections by about 20-30\%
 for \Lhyp{16}{N}. We can expect the effects are less important
 for lighter hypernuclei.

\section{Numerical Results}

 A large advantage of $(e,e'K^{+})$ is a variety of the kinematical
condition. We can select the four momentum of
the intermediate photon through the final electron($e'$) energy
and angle. To avoid large momentum transfers, which prevent
hyperons from sticking on nuclei, we should select suitable
kinematical conditions.
\par
 As an example, we show the kinematics for
 $^{16}$O$(e,e'K^{+})$\Lhyp{16}{N} in Fig.2.  Fig.2a shows
 that if final electrons($e'$) are detected at a finite
 angle($\theta{e'}>0^{\circ}, \phi_{e'}=0^{\circ}$), the angle
 for outgoing $K$ minimizing the momentum transfer($p_{B}$)
 is given by $\theta_{K}\sim\theta_{e'}, \phi_{K}=180^{\circ}$.
 Further, from Fig.2b we find that in the finite-angle case
 ($\theta_{K}\sim\theta_{e'}>0$) as
 mentioned above, there is  a final electron energy minimizing
 $p_{B}$, which is around 1.2GeV in the case of Fig.2b.  Our
 calculations are performed near such suitable kinematical conditions.
\par
 To examine what kinds of hypernuclear states are favorably excited
 in the reactions, we calculate the production cross sections for
 various hypernuclear states under a common kinematical condition,
 that is,
\begin{eqnarray}
   p_{e} & = & 3.0 {\rm GeV/c},   \nonumber \\
   p_{e'} & = & 1.2 {\rm GeV/c},  \\
   \theta_{e'} & = & 6^{\circ} ,\; \phi_{e'} = 0^{\circ},  \nonumber \\
   \theta_{K} & = & 10^{\circ} ,\; \phi_{K} = 180^{\circ}, \nonumber
\end{eqnarray}
in the laboratory frame.
\par
  The results are given in Tables 5-9. From these tables, we apparently
 find that the states with the largest $J_{f}$ for given nucleon-hole
 and hyperon states, that is, the stretched states are strongly
 excited. This can be easily understood by large momentum transfers
 in the reactions. However, the same reason makes absolute
 cross sections small. For \Lhyp{16}{N} and \Lhyp{12}{B} cases, the
 $[(p_{3/2})^{-1}_{p},(p_{3/2})_{\Lambda}]_{J_{f}=3}$ states
 are most favorably excited with cross sections 2.485 and 2.667
 nb/sr$^{2}$/GeV, respectively. These values correspond to about 1.3\%
 in the sticking probability, which is the ratio to
 $(2j+1)\times$(cross section of the elementary process).
 For \Lhyp{4}{H}$^{*}$, the cross
 section(2.260nb/sr$^{2}$/GeV) is much larger than those for the
 $[(s_{1/2})^{-1}_{p},(s_{1/2})_{\Lambda}]_{J_{f}=1}$ states in
 \Lhyp{16}{N} and \Lhyp{12}{B}. The reason is that recoils
 of light hypernuclei reduce effectively the momentum transfers in
 the hypernuclear center-of-mass frame(by $(A-1)/A$).
\par
  Similar behaviors can be seen for $\Sigma$ hypernuclear productions.
 $\Sigma$ hypernuclear states with $T_{f}=3/2$ can be excited. But,
 their cross sections are much smaller than those with $T_{f}=1/2$ by
 about two orders of magnitude.  Therefore, we show results
 only for $T_{f}=1/2$.
\par
  Another prominent feature is large cross sections for unnatural
 parity states. This is due to the dominant spin-flip amplitudes in
 the elementary processes $N(\gamma,K^{+})Y$. In fact,
 cross sections for \Lhyp{4}{H}$^{*}$ and \Shyp{4}{H}$^{*}$
 are larger than those for \Lhyp{4}{H} and \Shyp{4}{H} by
 two orders of magnitude, respectively. At more forward angles(
 $\theta_{K}<10^{\circ}$), these ratios become larger, as shown Fig.5.
 Such a behavior is one of typical characters of spin-flip processes.
 Small cross sections
 for $[(p_{3/2})^{-1}_{p},(p_{3/2})_{\Lambda}]_{J_{f}=2}$ states
 of \Lhyp{16}{N} and \Lhyp{12}{B} are also an evidence of
 the spin-flip nature.
\par
  In Tables 7-9, we give results of $\Sigma$ hypernuclear productions
 for two kinds of the elementary process, that is, PS and PS+delta
 models. Bennhold\cite{Be2} noted that PS+delta model give
 a little better $\chi^{2}$ than PS model for known data but
 the final conclusion would have to await further data. The
 differences between them, which are 30\%-50\%, may be regarded as
 a scale of uncertainties.
\par
 Angular distributions are shown in Figs.3-5. We can see that except
 for $A$=4 cases(Fig.5), the cross sections reduce rapidly for
 larger angles than $10^{\circ}$. In the present kinematics,
 $\theta_{K}\sim4^{\circ}$ gives the minimum momentum
 transfer$\sim$270MeV/c and $\theta_{K}=10^{\circ}$(15$^{\circ}$)
 does $\sim$320(420)MeV/C. In other word, for $\theta_{K}>10^{\circ}$,
 momentum transfers go rapidly away from the Fermi momentum
 ($\sim$280MeV/c).
\par
 For $A$=4 cases, a factor $(A-1)/A$ mentioned above makes the
 reduction mild, as shown in Fig.5. A peculiar behavior for
 $^{4}$H$(e,e'K^{+})$\Lhyp{4}{H}(\Shyp{4}{H}) comes from the
 dominant spin-flip ammplitudes in the elementary reactions.
\begin{table}
 \caption{Production cross sections for various states of
          \Lhyp{16}{N} in nb/sr$^2$/GeV}
 \begin{center}
 \begin{tabular}{cccc|cccc}
 \hline
 hole & $\Lambda$ & $J_{f}$ & & hole & $\Lambda$ & $J_{f}$ & \\ \hline
 $s_{1/2}$ & $s_{1/2}$ & 0 & 0.002 & $p_{3/2}$ & $p_{3/2}$ & 2 & 0.013 \\
           &           & 1 & 0.393 &           &           & 3 & 2.485 \\
           & $p_{3/2}$ & 1 & 0.203 &           & $p_{1/2}$ & 1 & 0.021 \\
           &           & 2 & 0.631 &           &           & 2 & 1.329 \\
           & $p_{1/2}$ & 0 & 0.002 & $p_{1/2}$ & $s_{1/2}$ & 0 & 0.033 \\
           &           & 1 & 0.369 &           &           & 1 & 0.844 \\
 $p_{3/2}$ & $s_{1/2}$ & 1 & 0.526 &           & $p_{3/2}$ & 1 & 0.023 \\
           &           & 2 & 1.576 &           &           & 2 & 1.320 \\
           & $p_{3/2}$ & 0 & 0.000 &           & $p_{1/2}$ & 0 & 0.002 \\
           &           & 1 & 0.388 &           &           & 1 & 0.555 \\
 \hline
 \end{tabular}
 \end{center}
\end{table}
\begin{table}
 \caption{Production cross sections for various states of
           \Lhyp{12}{B} in nb/sr$^2$/GeV}
 \begin{center}
 \begin{tabular}{cccc|cccc}
 \hline
 hole & $\Lambda$ & $J_{f}$ & & hole & $\Lambda$ & $J_{f}$ & \\ \hline
 $s_{1/2}$ & $s_{1/2}$ & 0 & 0.003 & $p_{3/2}$ & $p_{3/2}$ & 0 & 0.000 \\
           &           & 1 & 0.644 &           &           & 1 & 0.390 \\
           & $p_{3/2}$ & 1 & 0.223 &           &           & 2 & 0.020 \\
           &           & 2 & 0.709 &           &           & 3 & 2.667 \\
           & $p_{1/2}$ & 0 & 0.003 &           & $p_{1/2}$ & 1 & 0.024 \\
           &           & 1 & 0.354 &           &           & 2 & 1.252 \\
 $p_{3/2}$ & $s_{1/2}$ & 1 & 0.724 &           &           &   &  \\
           &           & 2 & 2.185 &           &           &   &  \\
 \hline
 \end{tabular}
 \end{center}
\end{table}
\begin{table}
 \caption{Production cross sections for various states of
          \Shyp{16}{N}($T_{f}=1/2$) in nb/sr$^2$/GeV}
 \begin{center}
 \begin{tabular}{ccccc}
 \hline
 hole & $\Sigma$ & $J_{f}$ & PS+delta & PS \\ \hline
 $s_{1/2}$ & $s_{1/2}$ & 0 & 0.001 & 0.001 \\
           &           & 1 & 0.013 & 0.005 \\
 $p_{3/2}$ &           & 1 & 0.077 & 0.055 \\
           &           & 2 & 0.314 & 0.176 \\
 $p_{1/2}$ &           & 0 & 0.055 & 0.021 \\
           &           & 1 & 0.093 & 0.056 \\ \hline
 \end{tabular}
 \end{center}
\end{table}
\begin{table}
 \caption{Production cross sections for various states of
          \Shyp{12}{B}($T_{f}=1/2$) in nb/sr$^2$/GeV}
 \begin{center}
 \begin{tabular}{ccccc}
 \hline
 hole & $\Sigma$ & $J_{f}$ & PS+delta & PS \\ \hline
 $s_{1/2}$ & $s_{1/2}$ & 0 & 0.001 & 0.001 \\
           &           & 1 & 0.029 & 0.018 \\
 $p_{3/2}$ &           & 1 & 0.112 & 0.085 \\
           &           & 2 & 0.420 & 0.276 \\ \hline
 \end{tabular}
 \end{center}
\end{table}
\begin{table}
 \caption{Production cross sections for \Yhyp{4}{H}
          and \Yhyp{4}{H}$^{*}$ in nb/sr$^2$/GeV}
 \begin{center}
 \begin{tabular}{ccc|cccc}
 \hline
    & $J_{f}$ & & & $J_{f}$ & PS+delta & PS \\ \hline
 $^{4}_{\Lambda}$H &0 &0.020 &$^{4}_{\Sigma}$H &0 &0.054 &0.077 \\
 $^{4}_{\Lambda}$H$^{*}$ & 1 & 2.260 &
 $^{4}_{\Sigma}$H$^{*}$ & 1 & 4.021 & 2.600 \\ \hline
 \end{tabular}
 \end{center}
\end{table}

\section{Comparison with Quasifree Hyperon Production}
  In the experiments, quasifree hyperon productions are the largest
 background process for hypernuclear productions.
 To examine the feasibility of the experiments, we have to estimate
 this background process.
\par
  Quasifree hyperon production cross sections are given by
\begin{equation}
    \frac{d\sigma}{d\omega}=\alpha \sigma R(q,\omega),
\end{equation}
 where $\sigma$ is the free elementary cross section,
 $R(q,\omega)$ is so-called a quasifree response function and
 $\alpha$ is a kinematical factor near 1.
\par
  Based on the picture of knockout processes from light nucleus,
 $R(q,\omega)$ is given by
\begin{equation}
 R(q,\omega)= (2\pi)\int_{S} dp_{N} \frac{p_{N}}{q}\rho(p_{N})
      |\omega+M_{i}-(M_{f}^2+p_{N}^2)^{1/2}|,
\end{equation}
 where, $M_{i}$($M_{f}$) is rest mass of initial target(final residual)
 nucleus and $\rho(p_{N})$ is the nucleon momentum distribution in
 target nucleus. The region $S$ is defined by
\begin{eqnarray}
  S=\left\{p_{N}\left|p_{N}\geq0,\sqrt{m_{Y}^2+(p_{N}-q)^2}
    \right. \right. &\leq& \omega+M_{i}-\sqrt{M_{f}^2+p_{N}^2}
    \nonumber \\
  & & \left. \leq \sqrt{m_{Y}^2+(p_{N}+q)^2} \right\}. \nonumber
\end{eqnarray}
\par
  Using this formalism, we obtain the result shown in Fig.6
 for the case of $^{4}$He target. The result depends strongly on
 the widths of hypernuclei. If the hypothetical \Shyp{4}{H}$^{*}$
 exists, we can observe clearly its signal. However, the signals of more
 possible states, for example,
 $[(p_{3/2})_{N}^{-1},(s_{1/2})_{\Sigma}]_{J_{f}=2}$
 states in \Shyp{16}{N} and \Shyp{12}{B} may be not so
 clear, because of their small cross
 sections(0.3-0.4 nb/sr$^{2}$/GeV), which
 correspond to about 10\% of that for \Shyp{4}{H}$^{*}$(see filled
 peaks in Fig.6).
\par
  For $\Lambda$ hypernuclei, Fig.6 shows that the electroproduction
 experiment is promising. We can expect to observe two or three
 unnatural parity states for in \Lhyp{16}{N} and \Lhyp{12}{B}.

\section{Conclusion}
  In the present work, we estimated the electroproduction of
 hypernuclei. We find large cross sections for the stretched states
 and unnatural parity states.  For $\Lambda$ hypernuclei, they are
 a few nb/sr$^{2}$/GeV, which seems sufficiently measurable in spite
 of large backgrounds by the quasifree hyperon productions.
  For $\Sigma$ hypernuclei, we find the cross sections of several
 tenth of nb/sr$^{2}$/GeV for \Shyp{16}{N} and
 \Shyp{12}{B}. These signals may not be measured clearly in the
 experiments.
 In these calculations, however, all $p$-states of
 $\Sigma$ particle were assumed to be unbound. If the bound $p$-states
 exist, we get measurable cross sections(a few nb) for the stretched
 states including such a
 $p$-state $\Sigma$. If \Shyp{4}{H}$^{*}$ with spin=1, which is
 not bound theoretically\cite{Ha1}-\cite{Ha3},
 exists, it can be clearly observed in the electroproduction
 experiments.
\par
  The present calculations include several uncertainties.
 Especially, the elementary processes at the energy region considered
 here is little known. Their phenomenological data are strongly
 desired.

\vspace{1.0cm}
\par
\noindent {\em Acknowledgement.}
\vspace{0.5cm}
\par
 The author wishes to thank Professor T. Saito for motivating him to
 this work and valuable discussions.

\vspace{1.0cm}

\noindent{\em Figure captions.}
\vspace{0.5cm}
\par
 Fig.1 a: Model of $N(e,e'K^{+})Y$ reactions. b: Model of
 $N(\gamma,K^{+})Y$ vertex.\\
 ($Y=\Lambda,\Sigma$)
\vspace{0.5cm}
\par
 Fig.2 Momentum transfer to hypernuclei as a function of $K^{+}$ angle,
 $\theta_{K}$(a) and final electron momentum, $p_{e'}$(b) in the case
 of $^{16}$O$(e,e'K^{+})^{16}_{\Lambda}$N($p_{e}$=\\
 3.0GeV/c, $\theta_{e'}=10^{\circ}$, $\phi_{e'}=0^{\circ}$).
 a: Solid and dotted lines are the cases with $\phi_{K}=0^{\circ}$
 and 180$^{\circ}$, respectively($p_{e'}$=1.2GeV/c).
 b: Solid and dotted lines are the cases with
 $\theta_{K}=0^{\circ}$ and 10$^{\circ}$, respectively
 ($\phi_{K}=180^{\circ}$).
\vspace{0.5cm}
\par
 Fig.3 Angular distributions of productions of $\Lambda$ hypernuclear
 stretched states in $^{16}$O$(e,e'K^{+})$\Lhyp{16}{N}(a,b) and
 $^{12}$C$(e,e'K^{+})$\Lhyp{12}{B}(c). Solid, dotted and dashed
 lines are those for
 hypernuclear states with the $s_{1/2}$, $p_{1/2}$ and $p_{3/2}$
 $\Lambda$ states, respectively.  Lines in a(upper and lower lines
 in b for each type of line) are those with the
 $s_{1/2}$($p_{3/2}$ and $p_{1/2}$) nucleon hole. Lower(upper) three
 lines in c are those with the $s_{1/2}$($p_{3/2}$)
 nucleon hole. $p_{e}$=3.0GeV/c,
 $p_{e'}$=1.2GeV/c, $\theta_{e'}=6^{\circ}$, $\phi_{e'}=0^{\circ}$,
 $\phi_{K}=180^{\circ}$.
\vspace{0.5cm}
\par
 Fig.4 The same as Fig.3 but in $^{16}$O$(e,e'K^{+})$\Shyp{16}{N}
 and $^{12}$C$(e,e'K^{+})$\Shyp{12}{B}. In all
 cases, $\Sigma$ states are $s_{1/2}$.  Solid, dotted, dashed and
 dash-dotted lines are those with the $s_{1/2}(J_{f}=1)$,
 $p_{3/2}(J_{f}=1)$,
 $p_{3/2}(J_{f}=2)$ and $p_{1/2}(J_{f}=1)$ nucleon hole, respectively.
\vspace{0.5cm}
\par
 Fig.5 The same as Fig.3 but in $^{4}$He$(e,e'K^{+})$\Lhyp{4}{H},
 \Lhyp{4}{H}$^{*}$, \Shyp{4}{H} and \Shyp{4}{H}$^{*}$,
 which correspond to lower solid, upper solid, lower dotted and upper
 dotted lines, respectively.
\vspace{0.5cm}
\par
 Fig.6 Sum of quasifree hyperon productions and hypernuclear
 productions in $^{4}$He$(e,e'K^{+})X$ at $\theta_{K}=10^{\circ}$(a)
 and $5^{\circ}$(b)($\phi_{K}=180^{\circ}$). Peaks labeled $\Lambda$
 and $\Sigma$ are \Lhyp{4}{H}$^{*}$($\Gamma$=4MeV) and
 \Shyp{4}{H}$^{*}$($\Gamma$=5MeV and 10MeV) productions,
 respectively. Filled peaks correspond to 10\% of
 \Shyp{4}{H}$^{*}$($\Gamma$=5MeV) peaks(shifted to the
 $B_{\Sigma}$=5.1MeV for \Shyp{16}{N}).


\begin{thebibliography}{99}
 \bibitem{Hsi} S. S. Hsiao and S. R. Cotanch,
               Phys. Rev. C28(1983), 1668.
 \bibitem{Ber} A. M. Bernstein, T. W. Donnelly and G. N. Epstein,
               Nucl. Phys. A358(1981), 195c.
 \bibitem{Ros} A. S. Rosenthal, D. Halderson, K. Hodgkinson and F. Tabakin,
               Ann. Phys. 184(1988), 33.
 \bibitem{Be1} C. Bennhold and L. E. Wright,
               Phys. Rev. C39(1989), 927.
 \bibitem{Tan} H. Tanabe, M. Kohno and C. Bennhold,
               Phys. Rev. C39(1989), 741.
 \bibitem{Be2} C. Bennhold,
               Phys. Rev. C39(1989), 1944.
 \bibitem{Be3} C. Bennhold,
               Nucl. Phys. A547(1992), 79c.
 \bibitem{Sot} M. Sotona, K. Itonaga, T. Motoba, O. Richter and J. \v{Z}ofka,
               Nucl. Phys. A547(1992), 63c.
 \bibitem{Coh} J. Cohen,
               Phys. Rev. C32(1985), 543.
 \bibitem{Tho} H. Thom,
               Phys. Rev. 151(1966), 1322.
 \bibitem{Ad1} R. A. Adelseck and L. E. Wright,
               Phys. Rev. C38(1988), 1965.
 \bibitem{Ad2} R. A. Adelseck, C. Bennhold and L. E. Wright,
               Phys. Rev. C32(1985), 1681.
 \bibitem{Kad} H. Kadowaki et al.,
               Prog. Theor. Phys. 79(1988), 263.
 \bibitem{Ad3} R. A. Adelseck and B. Saghai,
               Phys. Rev. C42(1990), 108.
 \bibitem{Hay} R. S. Hayano et al.,
               Phys. Lett. B231(1989), 355.
 \bibitem{Ha1} T. Harada, S. Shinmura, Y. Akaishi and H. Tanaka,
               Soryushiron-Kenkyu 76(1987), 25.
 \bibitem{Ha2} T. Harada, Y. Akaishi, S. Shinmura and H. Tanaka,
               Nuovo Cimento 102A(1989), 473.
 \bibitem{Ha3} T. Harada, S. Shinmura, Y. Akaishi and H. Tanaka,
               Nucl. Phys. A501 (1990), 715.
 \end{thebibliography}
\end{document}